\documentclass[aip,reprint,amsmath,amssymb,floatfix]{revtex4-2}

\usepackage[T1]{fontenc}
\usepackage[utf8]{inputenc}
\usepackage{graphicx}
\usepackage{bm}
\usepackage{mathptmx}
\usepackage{xcolor}
\usepackage{hyperref}
\newcommand{\rev}[1]{#1}
\newcommand{\newrev}[1]{#1}
\newcommand{\abstractrev}[1]{#1}
\newcommand{\currentrev}[1]{#1}
\hypersetup{hidelinks}
\usepackage{placeins}


\graphicspath{{figures/}}

\makeatletter
\let\auto@bib\@empty
\let\auto@bib@innerbib\@empty
\makeatother

\begin{document}

\preprint{Working manuscript -- confirmed second version, 2026-09-04}
\title{High-charge, highly polarized positron beams generated from a laser-driven nanowire-array target}

\author{De-Sheng Zhang}
\affiliation{Key Laboratory of Beam Technology of the Ministry of Education, and School of Physics and Astronomy, Beijing Normal University, Beijing 100875, China}
\author{Cui-Wen Zhang}
\affiliation{Institute of Applied Physics and Computational Mathematics, Beijing 100088, China}
\author{Kun Xue}
\email{xuekun@xjtu.edu.cn}
\affiliation{Ministry of Education Key Laboratory for Nonequilibrium Synthesis and Modulation of Condensed Matter, Shaanxi Province Key Laboratory of Quantum Information and Quantum Optoelectronic Devices, School of Physics, Xi'an Jiaotong University, Xi'an 710049, China}
\author{Feng Wan}
\affiliation{Ministry of Education Key Laboratory for Nonequilibrium Synthesis and Modulation of Condensed Matter, Shaanxi Province Key Laboratory of Quantum Information and Quantum Optoelectronic Devices, School of Physics, Xi'an Jiaotong University, Xi'an 710049, China}
\author{Xue-Ren Hong}
\affiliation{\mbox{College of Physics and Electronic Engineering, Northwest Normal University, Lanzhou 730070, China}}
\author{Jian-Xing Li}
\email{jianxing@xjtu.edu.cn}
\affiliation{Ministry of Education Key Laboratory for Nonequilibrium Synthesis and Modulation of Condensed Matter, Shaanxi Province Key Laboratory of Quantum Information and Quantum Optoelectronic Devices, School of Physics, Xi'an Jiaotong University, Xi'an 710049, China}
\author{Bai-Song Xie}
\email{bsxie@bnu.edu.cn}
\affiliation{Key Laboratory of Beam Technology of the Ministry of Education, and School of Physics and Astronomy, Beijing Normal University, Beijing 100875, China}
\affiliation{\mbox{Institute of Radiation Technology, Beijing Academy of Science and Technology, Beijing 100875, China}}
\date{September 9, 2026}

\begin{abstract}
\abstractrev{The generation of high-charge, highly polarized positron beams in
the interaction of a linearly polarized laser pulse with a nanowire-array target
is investigated.} \abstractrev{Here, laser-driven} electrons
emit high-energy photons through \rev{nonlinear Compton scattering (NCS)}, which
\rev{subsequently produce electron--positron pairs through the nonlinear
Breit--Wheeler (NBW) process.} We model this interaction using two-dimensional
spin-resolved \rev{quantum electrodynamics particle-in-cell (QED-PIC)} simulations.
\abstractrev{At positron birth, the sign of $S_z$ is statistically correlated
with that of the local $B_z$. The spatiotemporal field structure arising from
the laser--nanowire interaction strengthens the correlation between the birth
spin sign and the direction of the subsequent transverse Lorentz impulse, thereby
limiting cancellation between opposite-spin contributions at a given angle.}
\abstractrev{The results show that the average polarization degree reaches $|\bar S_z|\approx0.46$,
and the positron charge satisfying $|\bar S_z|>0.3$ is approximately
$308\,\mathrm{nC}$.} \newrev{Parameter scans reveal that the high-polarization positron charge is maximized at intermediate target densities and nanowire periods.} Such a source
  could enable polarization-sensitive studies of strong-field QED and spin-dependent phenomena in high-energy and materials physics.
\end{abstract}

\maketitle
\section{Introduction}
\abstractrev{Relativistic polarized positron beams are important tools for
searches for new physics and studies of nucleon spin structure.}
In high-energy collider experiments, positron-beam polarization can control the
helicity composition of the initial state, thereby improving sensitivity to
specific interactions and reaction mechanisms~\cite{MoortgatPick2008}. In nuclear and condensed-matter
physics, polarized positrons also provide spin-sensitive probes of nucleon
structure and material magnetism~\cite{Airapetian2007,Gidley1982}. For these applications, the
polarization degree determines the strength of spin-dependent signals, whereas
the positron charge governs the attainable event rate and statistical
precision. A positron source of practical value must therefore provide both
high polarization and sufficient charge.

Conventionally, polarized positrons are \currentrev{mainly} produced through a two-step process.
Polarized photons are first generated using a high-energy electron beam and
are subsequently converted into polarized positrons in a target. An
inverse-Compton-based experiment achieved a polarization of approximately
73\% with $2\times10^{4}$ positrons per bunch~\cite{Omori2006}, while the E166
experiment measured a polarization above 80\% near 6~MeV~\cite{Alexander2008};
polarization transfer from an MeV electron beam has also produced positron
polarization up to 82\%~\cite{Abbott2016}.
These schemes can provide high polarization but rely on high-quality
high-energy electron beams and dedicated conversion systems, with limited
positron yield. With advances in ultraintense laser technology, \rev{NBW} pair
creation in laser--electron-beam collisions has
\currentrev{been proposed as} a route to producing polarized positron beams. A two-color laser field
\currentrev{is expected to produce} positrons with approximately 60\% polarization~\cite{Chen2019}.
Elliptically polarized fields
\currentrev{can separate} oppositely polarized components in angle, with the polarization
approaching 90\% in selected angular regions~\cite{Wan2020}. Helicity transfer
from polarized electron beams \currentrev{has been shown theoretically to yield} 40--65\% polarization and
$10^{5}$--$10^{6}$ positrons per bunch~\cite{Li2020}. These schemes remain
constrained by electron-beam flux, initial beam polarization in some
configurations, and stringent spatiotemporal synchronization~\cite{Xue2022}. Direct
laser--solid interactions avoid external beam synchronization and \currentrev{can generate}
dense positron populations through strong-field QED processes
~\cite{Ridgers2012,Nerush2011,Zhu2016,Wang2017,Gu2018,Gu2019}. The interaction of a linearly polarized laser pulse with a flat foil is
predicted to produce approximately 30~nC of positrons with polarization above
30\%~\cite{Song2022}. Using an oblique foil and a self-generated quasi-static
magnetic field, \currentrev{simulations predict an average polarization of} approximately 70\%,
with a yield exceeding 0.1~nC~\cite{Xue2023}. Three-dimensional simulations of
high-density electron-beam interaction with a solid target yielded 0.2~nC of
positrons with an average polarization of approximately 40\%~\cite{Zhu2024}.
More recently, a cone-channel target driven by a relativistic electron beam
\currentrev{could produce} approximately 0.63~nC with an average polarization near 60\%
~\cite{Cao2026}.
\currentrev{However, many of these schemes combine high polarization with
a comparatively small positron charge.} Therefore, how to
\currentrev{obtain highly polarized positrons with sufficient usable charge}
remains an open challenge.

In this paper, we propose a scheme for generating a high-charge, highly
polarized positron source from a linearly polarized laser-driven
nanowire-array target. We investigate the formation and evolution of positron
polarization using two-dimensional spin-resolved \rev{QED-PIC}
simulations. NBW pair creation preferentially produces positrons
with spins aligned with the local magnetic field, giving rise to local spin
polarization. Subsequent transport can mix opposite-spin populations at the same
emission angle, reducing the angle-resolved polarization. However, the
spatiotemporal electromagnetic fields generated by the laser--nanowire interaction
strengthen the correlation between initial spin signs and subsequent transverse
Lorentz impulse directions, reducing this mixing. The simulations show that the
positron charge satisfying $|\bar S_z|>0.3$ reaches approximately
$308\,\mathrm{nC}$, about an order of magnitude higher than that obtained with a
flat foil. Further investigation shows that the high-polarization
positron yield does not increase monotonically with the total yield. To generate
high-charge, highly polarized positron beams, a target electron density near
$200n_c$ and a nanowire period near $1.0\,\mu\mathrm{m}$ should be selected.

\section{Simulation Setup}
\begin{figure}[!t]
  \centering
  \includegraphics[width=\columnwidth]{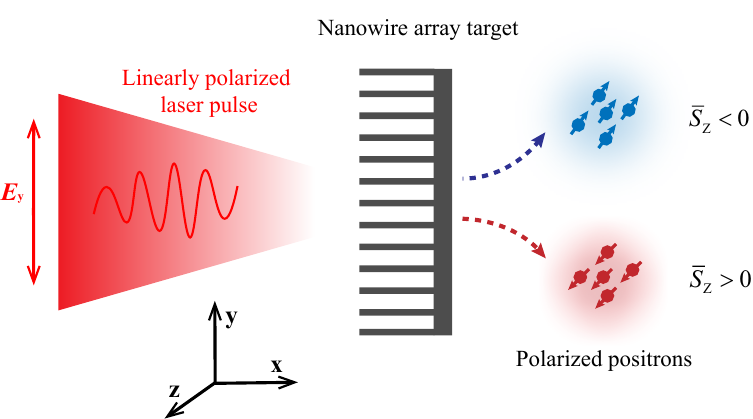}
  \caption{Schematic of polarized positron generation from a nanowire-array target. A linearly polarized laser pulse propagates along $+x$, with its electric field polarized along $y$, and interacts with the nanowire-array target. Relativistic electrons emit high-energy photons through \rev{NCS}, and these photons generate electron--positron pairs through the \rev{NBW} process. The resulting positrons are illustrated as two branches with opposite signs of the average spin component $\bar{S}_z$.}

  \label{fig:schematic}
\end{figure}

A linearly polarized laser pulse propagates along $+x$, with its electric field
polarized along $y$, and interacts with a nanowire-array target. The laser
drives and accelerates electrons from the nanowires. A fraction of these
electrons is driven against the laser-propagation direction and emits
high-energy photons through \rev{NCS}
~\cite{Ritus1985,DiPiazza2012}. Photons with
sufficiently large quantum parameter $\chi_\gamma$ subsequently generate
electron--positron pairs through \rev{NBW} pair creation
~\cite{Blackburn2017,DiPiazza2016}.
The resulting positrons form two spatially separated branches with opposite
transverse propagation directions and opposite signs of the average
polarization, as sketched in Fig.~\ref{fig:schematic}.

The simulations were performed with the two-dimensional spin- and
polarization-resolved QED-PIC code SLIPs~\cite{Wan2023}, which resolves the spin states of
electrons and positrons and the Stokes parameters of photons. \rev{NCS} and
\rev{NBW} pair creation are sampled using spin- and
polarization-resolved probabilities within the local constant-field
approximation~\cite{Duclous2011,Elkina2011,Ridgers2014,Gonoskov2015,SeiptKing2020}.
Photon propagation, pair creation, and subsequent particle
evolution are treated self-consistently. The reference simulation domain spans
$x\in[-20,40]~\mu\mathrm{m}$ and $y\in[-20,20]~\mu\mathrm{m}$, and is
discretized into $1800\times1200$ cells. The laser wavelength is
$\lambda_0=1~\mu\mathrm{m}$, with $T_0=\lambda_0/c$, and its normalized
amplitude is $a_0=eE_0/(m_e\omega_0c)=1500$, corresponding to a peak intensity
of approximately $3.08\times10^{24}~\mathrm{W\,cm^{-2}}$. The transverse
focal radius is $w_0=3~\mu\mathrm{m}$. The temporal profile is a Gaussian
  centered at $12T_0$ with a width parameter of $6T_0$. \rev{The laser is injected through the $x_{\min}$ boundary, with simple-outflow conditions applied at $x_{\max}$ and at both transverse boundaries.}

Aligned nanowire arrays can enhance laser-energy coupling, field penetration,
and electron acceleration relative to planar targets
~\cite{Purvis2013,Kaymak2016,Bargsten2017,Moreau2020,Fedeli2018,Ong2021}.
The target consists of fully ionized carbon plasma with an initially
unpolarized electron population. The nanowire array has an electron density
of $200n_c$, a transverse width of $0.2~\mu\mathrm{m}$, a period of
$\ell=0.8~\mu\mathrm{m}$, and a length of $10~\mu\mathrm{m}$ along $x$. The
patterned region spans $-10<y<10~\mu\mathrm{m}$ and is followed by a
$1~\mu\mathrm{m}$-thick solid substrate, where
$n_c=m_e\varepsilon_0\omega_0^2/e^2\simeq1.1\times10^{21}~\mathrm{cm^{-3}}$
is the critical density. The simulations use 200 electron and 40 carbon-ion
macroparticles per cell.

\section{Results and Discussion}

\begin{figure*}[!t]
  \centering
  \includegraphics[width=0.88\textwidth]{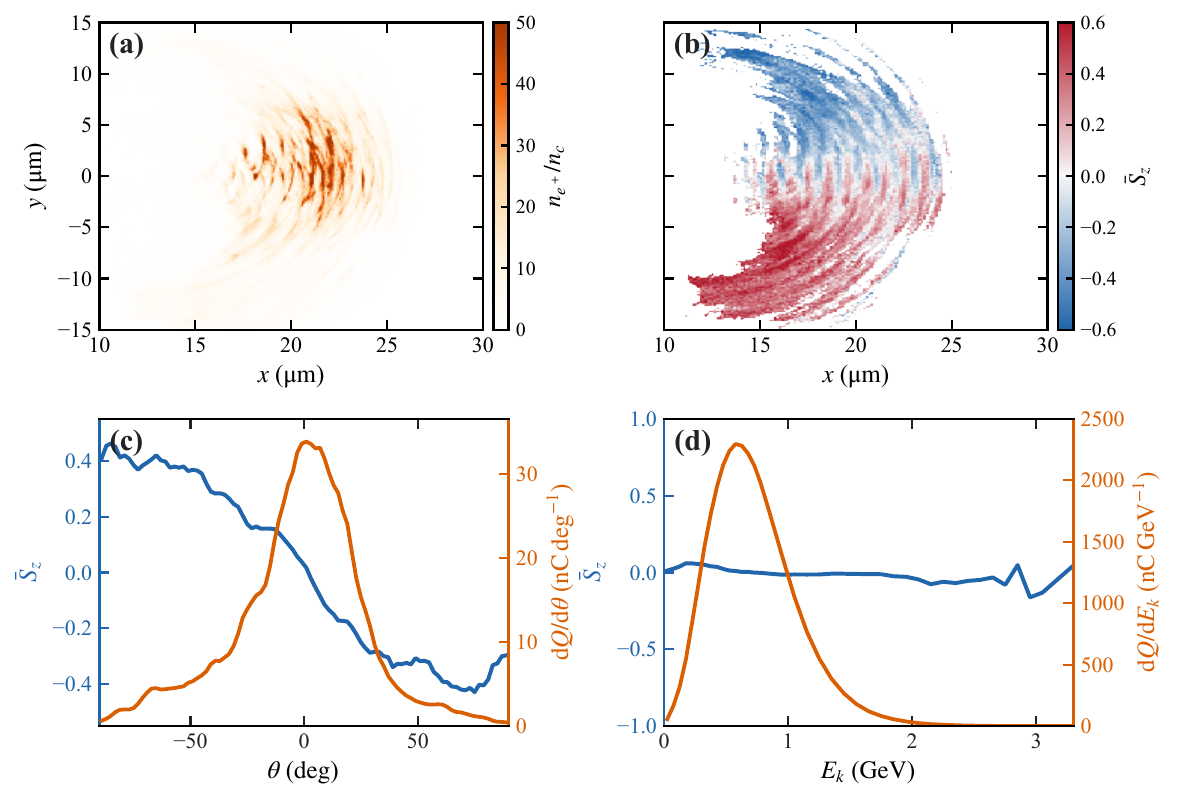}
  \caption{Spatial, angular, and energy distributions of the polarized positron beam. All distributions are evaluated at $t=55T_0$. (a,b) Spatial distributions of the positron density $n_{e^+}/n_c$ and average positron polarization $\bar{S}_z$ in the $x$--$y$ plane, respectively. (c) Positron angular distribution $\mathrm{d}Q/\mathrm{d}\theta$ (orange, right axis) and polarization $\bar{S}_z$ (blue, left axis) versus $\theta$, where $\theta=\arctan(p_y/p_x)$. (d) Positron energy spectrum $\mathrm{d}Q/\mathrm{d}E_k$ (orange, right axis) and polarization $\bar{S}_z$ (blue, left axis) versus the kinetic energy $E_k$.}

  \label{fig:beam}
\end{figure*}

Figures~\ref{fig:beam}(a) and~\ref{fig:beam}(b) present the spatial
distributions of the positron density $n_{e^+}$ and the average
polarization degree $\bar{S}_z$, respectively. The positrons form several
curved density bands extending mainly along the laser-propagation direction.
The different widths and density levels of these bands show that the source
has a clear spatial pattern but is not perfectly uniform. This may be related
to the laser interacting with successive nanowires while the laser intensity
and local plasma conditions are evolving, causing the positron yield to vary
from one interaction region to the next. The local positron density reaches
$n_{e^+}/n_c=89.2$, demonstrating that a high-density positron source can be
formed within a compact region. Figure~\ref{fig:beam}(b) further reveals two
spatially separated regions with opposite signs of the average polarization
degree: $\bar{S}_z$ is predominantly negative in the upper region and positive
in the lower region, with local values reaching approximately $\pm0.78$.
This separation shows that the average polarization is not uniformly mixed
across the beam. In principle, spatial apertures or magnetic transport could
be used to select different polarization branches, providing a possible route
towards branch-selective control of a polarized positron source.

Figure~\ref{fig:beam}(c) shows the positron charge distribution
$\mathrm{d}Q/\mathrm{d}\theta$ and the average polarization
degree $\bar{S}_z$ as functions of the emission angle $\theta$. The average
polarization degree is predominantly negative for $\theta>0$ and predominantly
positive for $\theta<0$. This angular separation is consistent with the
opposite-polarization regions in the upper and lower parts of the spatial
distribution shown in Fig.~\ref{fig:beam}(b). The maximum value of
$|\bar{S}_z|$ reaches approximately 0.46, exceeding the average polarization
of approximately 0.30 reported by Song~\textit{et al.} for a foil
target by about 50\%. This comparison highlights the stronger angular
polarization obtained with the nanowire-array target.
To identify a clearly polarized component, we consider positrons satisfying
$|\bar{S}_z|>0.3$. Experimentally, a 30\% beam polarization can produce a
measurable net spin asymmetry and improve the signal-to-background ratio in
polarization-sensitive measurements, making it a useful reference level for
this selection~\cite{MoortgatPick2008,Abbott2016}. The positron charge satisfying $|\bar{S}_z|>0.3$ reaches
approximately $\rev{308}\,\mathrm{nC}$, compared with approximately
$30\,\mathrm{nC}$ reported for the foil-target scheme, representing a
substantial increase of nearly one order of magnitude. This combination of
high charge and strong polarization is advantageous for experiments requiring
both a sizeable event rate and spin sensitivity, including spin-dependent
radiation and scattering measurements and positronium studies~\cite{Cassidy2018}.

Figure~\ref{fig:beam}(d) presents the positron charge distribution
$\mathrm{d}Q/\mathrm{d}E_k$ and the corresponding energy-resolved average
polarization degree as functions of the kinetic energy $E_k$. The positron
charge is concentrated near $0.6\,\mathrm{GeV}$ and extends to higher
energies. The average polarization remains close to zero over most of the
energy range and does not increase systematically with kinetic energy. The
energy spectrum therefore mainly describes how the positron charge is
distributed over kinetic energy, while the angular distribution provides the
more direct basis for selecting highly polarized positrons.

\begin{figure*}[!t]
  \centering
  \includegraphics[width=0.88\textwidth]{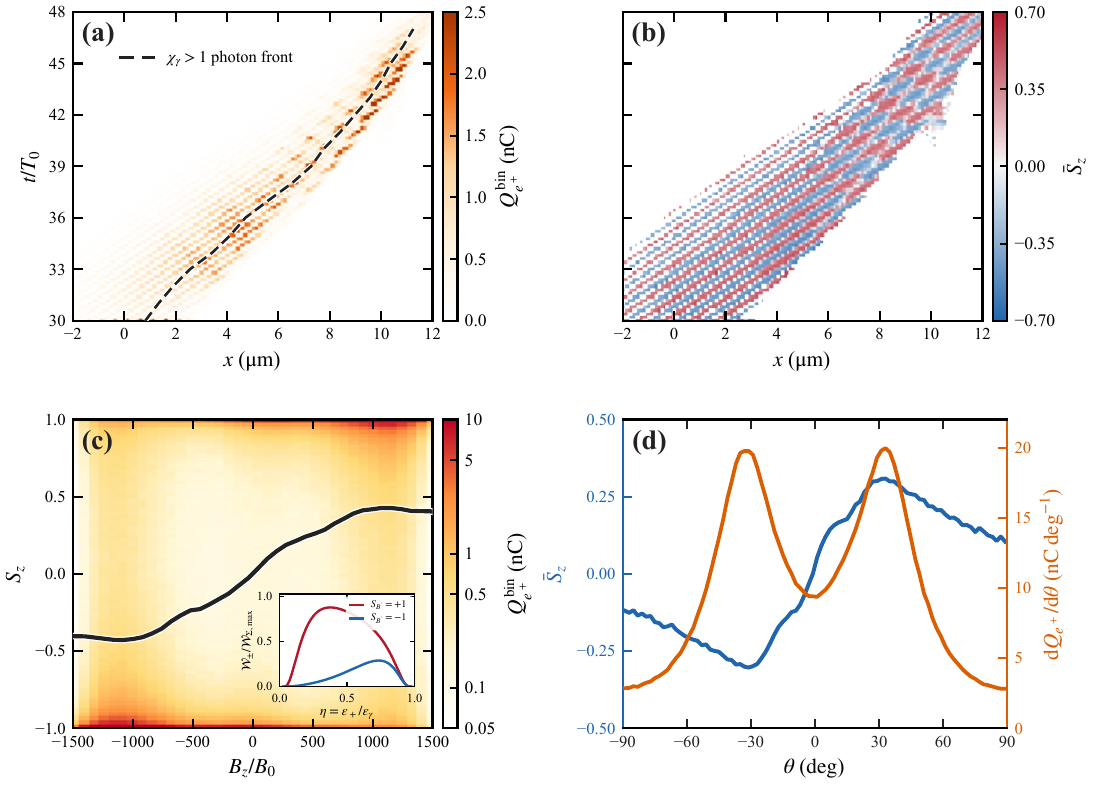}
  \caption{Spatiotemporal formation of the positron beam and its initial spin polarization. (a,b) Spatiotemporal distributions of the positron charge $Q_{e^+}^{\mathrm{bin}}$ and average polarization $\bar{S}_z$ in the $x$--$t$ plane, respectively. The black dashed curve in (a) shows the temporal evolution of the mean longitudinal position of photons with $\chi_\gamma>1$. (c) Positron charge distribution in the $B_z/B_0$--$S_z$ space at birth; the black curve shows the average polarization $\bar{S}_z$ versus $B_z/B_0$. The inset presents the spin-resolved \rev{NBW} rates $\mathcal{W}_{\pm}/\mathcal{W}_{\Sigma,\max}$ versus $\eta=\varepsilon_+/\varepsilon_\gamma$ for an unpolarized parent photon with $\chi_\gamma=2$, where $S_{B'}=+1$ ($-1$) denotes the positron spin parallel (antiparallel) to the magnetic field $\mathbf{B}'$ in the positron rest frame. (d) Positron angular distribution $\mathrm{d}Q_{e^+}/\mathrm{d}\theta$ (orange, right axis) and average polarization $\bar{S}_z$ (blue, left axis) at birth.}

  \label{fig:birth}
\end{figure*}

\rev{The phase-space coordinates and spin properties associated
with each positron are examined in Fig.~\ref{fig:birth} at the first recorded
instant following its creation.} Figures~\ref{fig:birth}(a) and
\ref{fig:birth}(b) show the positron charge $Q_{e^+}^{\mathrm{bin}}$ and
average polarization degree $\bar{S}_z$ in the $x$--$t$ plane, respectively.
In Fig.~\ref{fig:birth}(a), the positron-generation region forms a series of
diagonal bands. The dashed curve marks the mean longitudinal position of
photons with $\chi_\gamma>1$, for which \rev{NBW} pair creation
is strongly enhanced. The close agreement between this curve and the positron
bands shows that the generation region moves with the high-$\chi_\gamma$
photon front. Initially, the local field retains a clear phase structure
associated with the propagating laser pulse, so the phase-dependent
pair-production rate gives rise to a relatively regular sequence of bands. The
charge varies between these bands, showing that the positron yield changes
with propagation distance and time. Figure~\ref{fig:birth}(b) further reveals
alternating bands with positive and negative average polarization degrees. The
unequal charge levels in Fig.~\ref{fig:birth}(a) and the evolution of the
polarization bands in Fig.~\ref{fig:birth}(b) are both associated with the
plasma-induced modification of the local fields. As the laser pulse propagates
through the array, this field evolution alters the local positron yield and
the relative population of the two polarization branches, making the later
polarization bands broader and less distinct. This gradual evolution of the
band pattern indicates that the local field conditions vary continuously during
positron generation, rather than switching to a different mechanism
responsible for positron polarization. In the early stage, the local average
polarization degree reaches approximately $0.4$, demonstrating that a high
degree of polarization is already present at birth. Although contributions
from oppositely polarized bands can partially offset each other after spatial
integration, they provide the initial conditions for the subsequent selection
and transport of different polarization branches.

Figure~\ref{fig:birth}(c) directly compares the local magnetic-field component
$B_z$ with the $z$-component $S_z$ of the positron spin at the first recorded
instant following pair creation. For $B_z/B_0>0$, the average $S_z$ is
predominantly positive, with a mean value of approximately $+0.33$ in the
corresponding field range; for $B_z/B_0<0$, it is predominantly negative, with
a mean value of approximately $-0.33$. Regions where $B_z$ and $S_z$ have the
same sign contain about $68.5\%$ of the positron charge, indicating a clear
statistical correlation between the newborn positron spin and the local
magnetic-field direction.

\begin{figure*}[!t]
  \centering
  \includegraphics[width=0.96\textwidth]{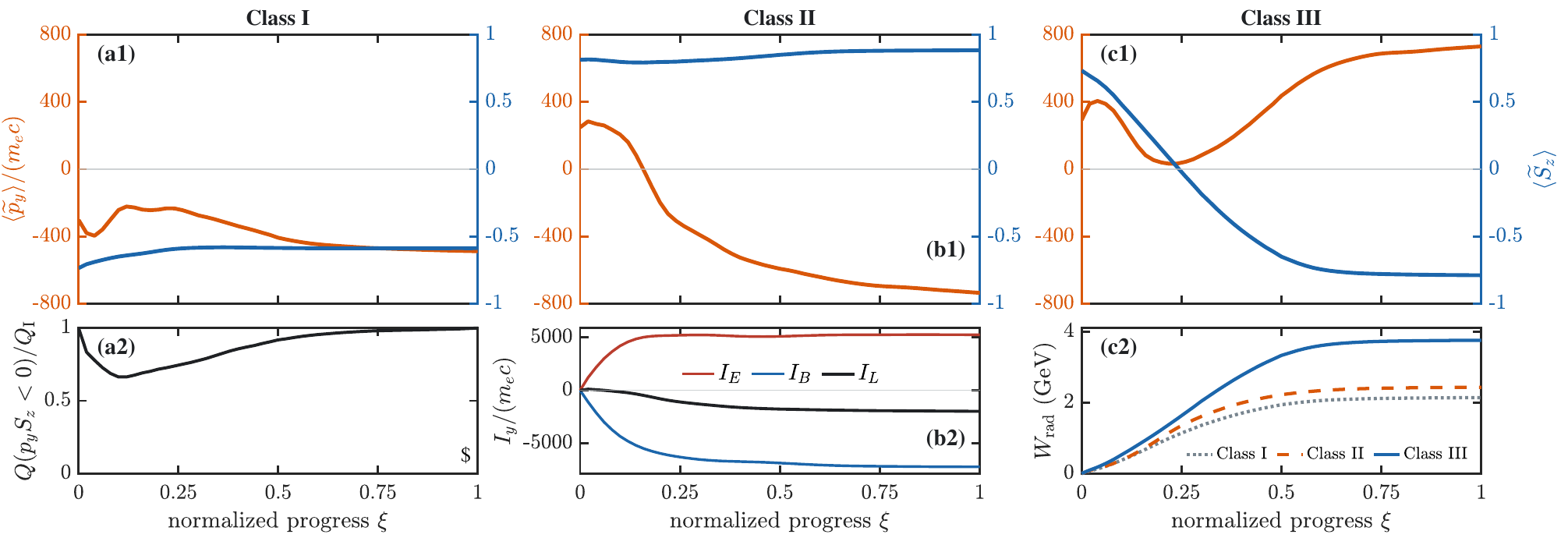}
  \caption{Spin dynamics of three positron classes. Positrons with $p_{y,f}S_{z,f}<0$ are divided into three classes according to the signs of $p_y$ and $S_z$ at production and at the final snapshot $t_f=55.0T_0$. Here, $p_{y,f}$ and $S_{z,f}$ denote the transverse-momentum component and $z$-component of spin in the final snapshot, respectively. (a1--c1) Evolution of the mean transverse momentum $\langle\widetilde p_y\rangle$ (orange, left axis) and mean spin component $\langle\widetilde S_z\rangle$ (blue, right axis). The corresponding variables are $\widetilde p_y=p_y\operatorname{sgn}(S_{z,b})$ and $\widetilde S_z=S_z\operatorname{sgn}(p_{y,b})$, where $S_{z,b}$ and $p_{y,b}$ are the spin component and transverse-momentum component at the production time $t_b$. The horizontal coordinate is $\xi=(t-t_b)/(t_f-t_b)$, with $\xi=0$ at production and $\xi=1$ at the final snapshot. (a1) In Class I, $p_y$ and $S_z$ have opposite signs both at production and in the final snapshot. (a2) Fraction of Class-I charge satisfying $p_yS_z<0$ during the evolution. (b1) In Class II, $p_y$ changes sign while $S_z$ retains its sign. (b2) Electric and magnetic contributions to the cumulative transverse impulse, $I_E$ and $I_B$, together with the total impulse $I_L$, for Class II. (c1) In Class III, $S_z$ changes sign while $p_y$ retains its sign. (c2) Estimated cumulative radiative energy loss $W_{\rm rad}$ for Classes I--III.}

  \label{fig:dynamics}
\end{figure*}

\begin{figure*}[!t]
  \centering
  \includegraphics[width=0.88\textwidth]{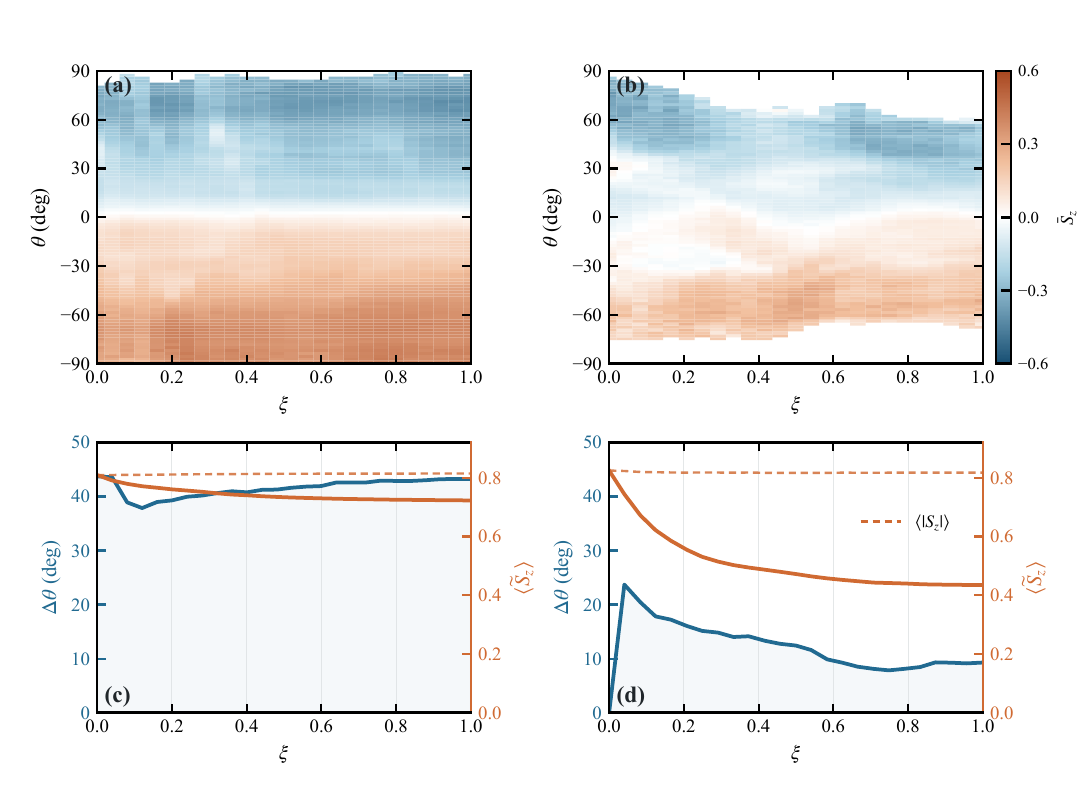}
  \caption{Evolution of angular polarization, angular separation and average spin. (a,b) Average positron polarization $\bar{S}_z$ versus $\theta$ and $\xi$ for all forward positrons from the nano target and flat foil, respectively. $\xi$ is normalized separately within the tracked windows $45.0T_0$--$50.0T_0$ for the nano target and $25.0T_0$--$30.0T_0$ for the flat foil; thus, $\xi=0$ denotes the beginning of each tracked window rather than positron birth. (c,d) Evolution of the angular separation $\Delta\theta$ (blue, left axis), the sign-aligned average polarization $\langle\widetilde{S}_z\rangle=\langle S_z\operatorname{sgn}(S_{z,0})\rangle$ (solid orange, right axis), and the average spin magnitude $\langle|S_z|\rangle$ (orange dashed, right axis) for positrons selected from the angular distribution at the start of each tracked time window for which the absolute average polarization satisfies $|\bar{S}_z(\theta)|>0.3$. The angular separation is defined as $\Delta\theta=|\langle\theta\rangle_{S_{z,0}>0}-\langle\theta\rangle_{S_{z,0}<0}|$.}

  \label{fig:transport}
\end{figure*}

To examine the microscopic origin of this correlation, the inset shows the
normalized spin-resolved \rev{NBW} rates
$\mathcal{W}_{\pm}/\mathcal{W}_{\Sigma,\max}$ as functions of the energy
fraction $\eta=\varepsilon_+/\varepsilon_\gamma$. Here,
$S_{B'}=+1$ and $S_{B'}=-1$ denote positron spin parallel and antiparallel,
respectively, to the local magnetic field $\mathbf{B}'$ in the positron rest
frame. For an unpolarized parent photon, with
$\xi'_1=\xi'_2=\xi'_3=0$, the spin-resolved pair-production rate reduces to
\cite{SeiptKing2020,Wan2023}
\begin{equation}
\frac{\mathrm d^2 W_{\mathrm{pairs}}}
{\mathrm d\varepsilon_+\,\mathrm dt}
=C_{\mathrm{pairs}}
\left[
D_0-\frac{1}{\eta}K_{1/3}(y)
\left(\mathbf{S}_+\cdot\mathbf{e}'_2\right)
\right],
\label{eq:fig3_spin_rate}
\end{equation}
where
\begin{equation}
\begin{aligned}
D_0&=\frac{\eta^2+(1-\eta)^2}{\eta(1-\eta)}K_{2/3}(y)
      +I_{1/3}(y),\\
I_{1/3}(y)&=\int_y^\infty K_{1/3}(u)\,\mathrm du,\\
\eta&=\frac{\varepsilon_+}{\varepsilon_\gamma},
\qquad y=\frac{2}{3\chi_\gamma\eta(1-\eta)}.
\end{aligned}
\label{eq:fig3_spin_definitions}
\end{equation}
Here, $\varepsilon_\gamma$ and $\varepsilon_+$ are the parent-photon and
positron energies, respectively, $\chi_\gamma$ is the photon quantum
parameter, $\mathbf{S}_+$ is the newborn positron spin vector,
$\mathbf{e}'_2$ is the local spin-basis vector used for the \rev{NBW process},
$K_{1/3}$ and $K_{2/3}$ are modified Bessel functions
of the second kind, and $C_{\mathrm{pairs}}$ is a spin-independent prefactor.
The inset uses $\chi_\gamma=2$.

The direction of $\mathbf{e}'_2$ should be interpreted together with the
local-field geometry. In the local QED basis convention used here and in the
high-energy, near-collinear limit,
\begin{equation}
\hat{\mathbf b}_+=\hat{\mathbf v}_+\times\hat{\mathbf a}_+
\approx-\hat{\mathbf k}_\gamma\times\hat{\mathbf E}'
\approx-\hat{\mathbf B}',
\label{eq:fig3_field_basis}
\end{equation}
and $\mathbf{e}'_2$ is oriented along $\hat{\mathbf b}_+$, giving
$\mathbf{e}'_2\approx-\hat{\mathbf B}'$. This approximate relation depends
on the chosen local QED basis and fixes the sign convention used for
interpreting the inset. Consequently, when the positron spin is oriented
along $\mathbf{B}'$, $\mathbf{S}_+\cdot\mathbf{e}'_2$ is negative, so that
the spin-dependent term enhances the rate; the opposite spin orientation
produces a suppressing contribution. Under the $\chi_\gamma=2$ conditions
used for the inset, the $S_{B'}=+1$ branch lies above the $S_{B'}=-1$ branch.
This provides microscopic support for the sign correlation between $B_z$ and
$S_z$ observed in the main panel, consistent with the magnetic-field-related
spin-quantization-axis analysis of \rev{NBW} pair production
~\cite{SeiptKing2020,Song2021,Dai2022,Zhao2023}.

Figure~\ref{fig:birth}(d) presents the birth-stage positron charge distribution,
$\mathrm{d}Q_{e^+}/\mathrm{d}\theta$, and the corresponding average polarization
degree $\bar{S}_z$ as functions of the emission angle $\theta$. The angular
distribution exhibits two approximately symmetric peaks near $\theta\approx
-30^\circ$ and $+30^\circ$. Each angular branch carries approximately
$9.2\times10^2\,\mathrm{nC}$, showing that both branches contain substantial
and comparable positron charge rather than arising from a small number of
large-angle particles. The average polarization degree is predominantly
negative for $\theta<0$ and predominantly positive for $\theta>0$, reaching
local values of approximately $-0.30$ and $+0.31$, respectively. These results
show that the angular branches already carry distinct polarization signatures
at the production stage.

This birth-stage pattern differs from the positron beam characterized after the
laser--plasma interaction in Fig.~\ref{fig:beam}(c). At birth, the charge is
concentrated in two off-axis angular branches rather than near $\theta\approx0$,
and the polarization signs on the two sides are reversed: the corresponding
beam is predominantly positively polarized for $\theta<0$ and negatively
polarized for $\theta>0$. These differences indicate that transport after pair
creation reorganizes the relation between emission angle and spin. Although the
average polarization degree of the entire birth-stage positron population is
close to zero, this mainly results from the two oppositely polarized angular
branches offsetting one another, rather than from an absence of local
polarization. Figure~\ref{fig:birth} therefore connects the magnetic-field-related
spin selection at pair creation with the formation of the initial
angle--polarization branches, providing the initial conditions for analyzing the
subsequent spin--momentum transport.

Starting from the birth-stage initial conditions identified in
Fig.~\ref{fig:birth}, we next examine how these angle--polarization branches
evolve during subsequent spin--momentum transport into the final distribution
shown in Fig.~\ref{fig:beam}(c). Figure~\ref{fig:beam}(c) shows that the
positron charge satisfying $|\bar{S}_z|>0.3$ reaches approximately
$\rev{308}\,\mathrm{nC}$. To determine how these positrons evolve after
birth, we select them at $t_f=55.0T_0$ and trace them back to the first
recorded instant following pair creation.

Among these positrons, those satisfying the target spin--angle relation
$p_{y,f}S_{z,f}<0$ carry approximately $215.5\,\mathrm{nC}$, corresponding to
about 70\% of the selected charge. Accordingly, the remaining approximately 30\% satisfy
$p_{y,f}S_{z,f}>0$ and have spin directions opposite to the dominant spin
orientation at their emission angles, thereby reducing the average polarization
degree. Figure~\ref{fig:dynamics} examines how the final relation
$p_{y,f}S_{z,f}<0$ is established and therefore takes positrons satisfying this
condition as the target branch. According to the endpoint signs of $p_y$ and
$S_z$ at birth and in the final snapshot, this target branch can be divided into
Classes I, II, and III, which account for 43.4\%, 30.8\%, and 25.8\% of the
total target-branch charge, respectively.

To compare the two branches at positive and negative angles using a common sign
convention, Figs.~\ref{fig:dynamics}(a1)--\ref{fig:dynamics}(c1) use the
transformed variables
\[
\widetilde p_y=p_y\operatorname{sgn}(S_{z,b}),\qquad
\widetilde S_z=S_z\operatorname{sgn}(p_{y,b}),
\]
where $p_{y,b}$ and $S_{z,b}$ are the transverse momentum and spin component at
birth, respectively. The sign functions use the birth-stage spin and
transverse-momentum signs of each particle as references to transform its
subsequent $p_y$ and $S_z$. This transformation preserves the absolute values
of the two quantities while mapping the two branches at positive and negative
angles onto a common sign convention, thereby preventing cancellation between
opposite branches when $p_y$ or $S_z$ is averaged. At birth, particles
satisfying $p_{y,b}S_{z,b}<0$ are mapped to $\widetilde p_y<0$ and
$\widetilde S_z<0$, whereas those satisfying $p_{y,b}S_{z,b}>0$ are mapped to
$\widetilde p_y>0$ and $\widetilde S_z>0$. Thus, when the particles within a
class share a net endpoint sign reversal in one of the two quantities, the
corresponding mean value also changes sign and crosses zero.

\begin{figure}[!t]
  \centering
  \includegraphics[width=0.78\columnwidth]{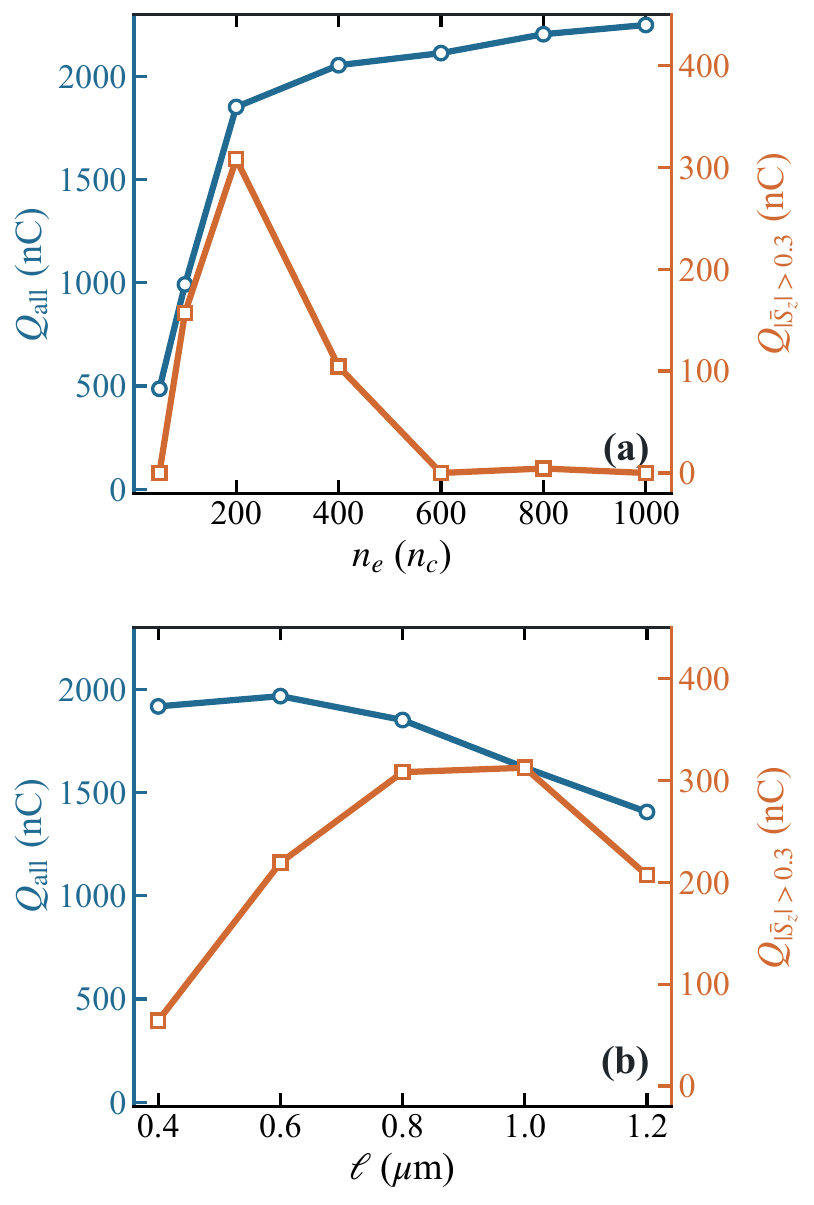}
  \caption{Parameter dependence of the total and high-polarization positron charge at $t=55T_0$. (a,b) Total forward positron charge $Q_{\rm all}$ (blue circles) and charge $Q_{|\bar{S}_z|>0.3}$ (orange squares) as functions of the electron density $n_e$ in units of $n_c$ and the nanowire period $\ell$, respectively. Here, $Q_{|\bar{S}_z|>0.3}$ denotes the charge of positrons selected from the angular distribution for which the absolute average polarization satisfies $|\bar{S}_z(\theta)|>0.3$.}

  \label{fig:robustness}
\end{figure}

Class I satisfies $p_yS_z<0$ both at birth and in the final snapshot,
representing a pathway in which the spin--angle relation established at birth
is retained during subsequent transport. In Fig.~\ref{fig:dynamics}(a1), both
$\langle\widetilde p_y\rangle$ and $\langle\widetilde S_z\rangle$ remain
negative throughout the evolution. Figure~\ref{fig:dynamics}(a2) further shows
that the Class-I charge fraction satisfying $p_yS_z<0$ decreases to a minimum
of approximately 0.66 at an early stage and then returns to 1. Thus, some
Class-I positrons temporarily do not satisfy $p_yS_z<0$ during transport, but
all recover this relation by the final snapshot.

At birth, Class II satisfies $p_yS_z>0$. As shown in
Fig.~\ref{fig:dynamics}(b1), $\langle\widetilde p_y\rangle$ changes from
positive to negative by crossing zero, whereas $\langle\widetilde S_z\rangle$
remains positive. Class II is therefore characterized mainly by a net sign
conversion of the transverse momentum, through which these positrons enter the
final target branch satisfying $p_{y,f}S_{z,f}<0$.
Figure~\ref{fig:dynamics}(b2) shows the cumulative transverse impulses
associated with this evolution. The electric contribution $I_E$ is positive,
the magnetic contribution $I_B$ is negative, and the total transverse Lorentz
impulse $I_L$ remains negative. Its direction is consistent with the net
transverse-momentum reversal shown in Fig.~\ref{fig:dynamics}(b1).

Class III also satisfies $p_yS_z>0$ at birth, but follows a different pathway.
Figure~\ref{fig:dynamics}(c1) shows that $\langle\widetilde p_y\rangle$ remains
positive, whereas $\langle\widetilde S_z\rangle$ crosses zero and becomes
negative. Class III is therefore characterized mainly by a net sign conversion
of the spin component, which brings the positrons into the target branch.
Figure~\ref{fig:dynamics}(c2) compares the cumulative radiative loss
$W_{\mathrm{rad}}$ for the three classes. The median cumulative radiative
losses are approximately $1.61$, $2.09$, and $3.65\,\mathrm{GeV}$ for Classes
I, II, and III, respectively, with Class III showing the largest loss. Since
Classes II and III both satisfy $p_yS_z>0$ at birth, their different
$W_{\mathrm{rad}}$ values provide a direct comparison between
transverse-momentum conversion and spin conversion. This comparison indicates
that the Class-III spin conversion is associated with stronger cumulative
radiative loss. Because $W_{\mathrm{rad}}$ is inferred from the difference
between the accumulated field work and the particle-energy change, it does not
identify individual photon-emission events.
Taken together, the three classes
show that the birth-stage polarization branches provide the initial conditions,
while subsequent spin--momentum transport determines how these branches are
retained or converted in the final state.

Figure~\ref{fig:transport} further examines why the nanowire target yields much more positron charge satisfying $|\bar S_z|>0.3$ than the flat foil. Panels (a,b) show the evolution of the angle-resolved average polarization for all forward positrons, whereas panels (c,d) show the evolution of the angular and spin diagnostics for positrons selected from angular regions satisfying $|\bar S_z(\theta)|>0.3$ at the start of each tracked time window. In the nanowire target, Fig.~\ref{fig:transport}(a) shows that the positive- and negative-polarization regions remain mainly separated into opposite angular regions throughout the evolution. Although the widths and strengths of the branches change, the correspondence between emission angle and polarization sign remains clear. By contrast, in the flat foil shown in Fig.~\ref{fig:transport}(b), the positive- and negative-polarization populations progressively mix, and their branch separation at the start of each tracked time window is weakened. When positrons with opposite spin signs occupy the same angular range, their contributions cancel, reducing $|\bar S_z|$. In the nanowire target, the preserved angular separation therefore allows a larger positron charge to remain in regions with $|\bar S_z|>0.3$.

Panels (c,d) track the angular separation and spin measures of the selected positron population. Here, $\Delta\theta$ denotes the separation between the angular centroids of the two groups with opposite spin signs defined at the start of each tracked time window. The solid curve shows $\langle\widetilde S_z\rangle=\langle S_z\operatorname{sgn}(S_{z,0})\rangle$, whereas the dashed curve shows the average spin magnitude $\langle|S_z|\rangle$. In the nanowire target, $\Delta\theta$ starts at approximately $44^\circ$, decreases briefly to about $38^\circ$, and then recovers to approximately $43^\circ$ by $\xi=1$. The solid $\langle\widetilde S_z\rangle$ curve remains high, reaching about $0.72$ at $\xi=1$, while $\langle|S_z|\rangle$ remains close to $0.81$. In the flat foil, $\Delta\theta$ briefly increases to about $24^\circ$ before decreasing to approximately $9^\circ$, and $\langle\widetilde S_z\rangle$ decreases from about $0.82$ to $0.43$. In contrast, $\langle|S_z|\rangle$ remains close to $0.82$. The similar average spin magnitudes in the two targets, together with the larger $\Delta\theta$ and larger $\langle\widetilde S_z\rangle$ in the nanowire target, show that its larger high-polarization positron charge does not primarily result from a larger spin magnitude. Instead, it results from the preservation of the correspondence between spin sign and emission angle, which reduces cancellation in the angle-resolved average.

Figure~\ref{fig:robustness} further examines how parameter variations affect the final high-polarization positron charge. The scans vary the normalized electron density $n_e/n_c$ and the nanowire period $\ell$, and compare the total forward positron charge $Q_{\rm all}$ with the charge $Q_{|\bar S_z|>0.3}$ selected from angular regions satisfying $|\bar S_z(\theta)|>0.3$. Figure~\ref{fig:robustness}(a) shows that $Q_{\rm all}$ increases from approximately $486~\mathrm{nC}$ at $n_e=50n_c$ to about $2249~\mathrm{nC}$ at $n_e=1000n_c$. However, $Q_{|\bar S_z|>0.3}$ does not increase with the total charge: it is zero at $50n_c$, reaches roughly $157~\mathrm{nC}$ at $100n_c$, reaches $308~\mathrm{nC}$ at $200n_c$, decreases to around $104~\mathrm{nC}$ at $400n_c$, and becomes nearly zero at the $600$--$1000n_c$ points. Thus, in this density scan, the largest total positron charge at high density does not coincide with the largest high-polarization charge. Instead, the largest $Q_{|\bar S_z|>0.3}$ occurs near $n_e=200n_c$. This indicates that the high-polarization positron charge depends not only on the total positron yield, but also on whether the angular distribution retains sufficient polarization contrast.

Figure~\ref{fig:robustness}(b) shows the effect of the nanowire period $\ell$. The total forward positron charge increases slightly from about $1918~\mathrm{nC}$ at $\ell=0.4~\mu\mathrm{m}$ to roughly $1967~\mathrm{nC}$ at $0.6~\mu\mathrm{m}$, and then decreases with increasing period to around $1406~\mathrm{nC}$ at $1.2~\mu\mathrm{m}$. By contrast, $Q_{|\bar S_z|>0.3}$ increases from approximately $64~\mathrm{nC}$ at $0.4~\mu\mathrm{m}$ to about $219~\mathrm{nC}$ at $0.6~\mu\mathrm{m}$, remains near $308$--$312~\mathrm{nC}$ for $\ell=0.8$--$1.0~\mu\mathrm{m}$, and then decreases to roughly $207~\mathrm{nC}$ at $1.2~\mu\mathrm{m}$. The period scan therefore also shows that the total forward charge is not sufficient to determine the high-polarization charge. Within the range examined here, intermediate nanowire periods are more favorable for producing a large high-polarization positron charge.

Taken together, the two scans show that $Q_{|\bar S_z|>0.3}$ responds differently to density and period than $Q_{\rm all}$, reaching large values only within finite parameter ranges rather than increasing directly with the total charge. This is consistent with the conclusion from Fig.~5 that the nanowire advantage is linked to the preservation of the correspondence between spin sign and emission angle during transport. This correspondence reduces cancellation between oppositely polarized contributions in the angle-resolved average, allowing more positrons to remain in angular regions with high average polarization. Figure~\ref{fig:robustness} therefore identifies the parameter ranges in which the nanowire target produces a large high-polarization positron charge.

\section{Conclusion}
A scheme for generating a high-charge, highly polarized positron source from a linearly polarized laser-driven nanowire-array target was investigated using two-dimensional spin-resolved QED-PIC simulations. \rev{NBW pair creation preferentially produces positrons with spins aligned with the local magnetic field, leading to local spin polarization. Subsequent transport can mix opposite-spin populations at the same emission angle and reduce the angle-resolved polarization. However, the spatiotemporal electromagnetic fields generated by the laser--nanowire interaction strengthen the correlation between initial spin signs and subsequent transverse Lorentz impulse directions, limiting this mixing.} The angle-resolved average polarization reaches $|\bar S_z|\approx0.46$. The positron charge satisfying $|\bar S_z|>0.3$ is approximately $\rev{308}\,\mathrm{nC}$, \newrev{an order of magnitude higher than} the corresponding $30\,\mathrm{nC}$ obtained with the flat foil\newrev{~\cite{Song2022}}.

The dependence of the total forward-positron charge and the high-polarization positron charge on the electron density $n_e$ and nanowire period $\ell$ was also examined. The largest high-polarization charge is obtained near an electron density of $200\,n_c$ and for nanowire periods \rev{near $1.0\,\mu\mathrm{m}$}. This result shows that increasing the total positron yield alone is insufficient to obtain a larger high-polarization yield; the angle-resolved polarization structure must also be preserved. Such a positron beam, combining hundred-nanocoulomb-level charge with a clear net polarization, could enhance spin-asymmetry signals in polarization-sensitive measurements while maintaining a high event rate, providing a potential positron source for spin-dependent radiation and scattering measurements and positronium studies.

\begin{acknowledgments}
This work was supported by the National Natural Science Foundation of China
(Grant Nos. 12375240, 12475249, 12535015, 12565023, and 12505235), the China
Postdoctoral Science Foundation (Grant No. 2024M762568), the Postdoctoral
Fellowship Program of CPSF (Grant No. GZC20252248), the Fundamental Research
Funds for the Central Universities of the Ministry of Education of China (Grant
No. xzy012025079), and the National Key Research and Development (R\&D) Program
(Grant No. 2024YFA1612700). The
computations were performed using computing resources at Xi'an Jiaotong
University and Beijing Normal University.
\end{acknowledgments}

\section*{Author Declarations}
\subsection*{Conflict of Interest}
The authors have no conflicts to disclose. 

\subsection*{Author Contributions}

\section*{Data Availability}
There are no publicly available research data or software supporting this
manuscript. Requests for further information or data should be sent to the
authors.

\end{document}